# Rotating-frame relaxation as a noise spectrum analyzer of a superconducting qubit undergoing driven evolution

Fei Yan[1], Simon Gustavsson[2], Jonas Bylander[2], Xiaoyue Jin[2], Fumiki Yoshihara[3], David G. Cory[1,4,5], Yasunobu Nakamura[3,6], Terry P. Orlando[2], and William D. Oliver[2,7]

[1]*Department of Nuclear Science and Engineering,*

*Massachusetts Institute of Technology (MIT), Cambridge, MA 02139, USA*

[2]*Research Laboratory of Electronics,*

*MIT, Cambridge, MA 02139, USA*

[3]*The Institute of Physical and Chemical Research*

*(RIKEN), Wako, Saitama 351-0198, Japan*

[4]*Institute for Quantum Computing and Department of Chemistry,*

*University of Waterloo, Waterloo, Ontario N2L 3G1, Canada*

[5]*Perimeter Institute for Theoretical Physics, Waterloo, ON, N2J 2W9, Canada*

[6]*Research Center for Advanced Science and Technology,*

*The University of Tokyo, Komaba, Meguro-ku, Tokyo 153-8904, Japan*

[7]*MIT Lincoln Laboratory, 244 Wood Street, Lexington, MA 02420, USA*

## Abstract

Gate operations in a quantum information processor are generally realized by tailoring specific periods of free and driven evolution of a quantum system. Unwanted environmental noise, which may in principle be distinct during these two periods, acts to decohere the system and increase the gate error rate. While there has been significant progress characterizing noise processes during free evolution, the corresponding driven-evolution case is more challenging as the noise being probed is also extant during the characterization protocol. Here we demonstrate the noise spectroscopy $(0.1 - 200\,\text{MHz})$ of a superconducting flux qubit during driven evolution by using a robust spin-locking pulse sequence to measure relaxation $(T_{1\rho})$ in the rotating frame. In the case of flux noise, we resolve spectral features due to coherent fluctuators, and further identify a signature of the $1\,\text{MHz}$ defect in a time-domain spin-echo experiment. The driven-evolution noise spectroscopy complements free-evolution methods, enabling the means to characterize and distinguish various noise processes relevant for universal quantum control.

**Introduction**

Within the gate-based model of quantum computation, a quantum algorithm may be efficiently implemented on a quantum information processor by decomposing it into a finite and discrete set of coherent gate operations [1]. Their realization, although in detail dependent on the particular physical-qubit modality, may be broadly divided into two categories, free-evolution and driven-evolution, indicating the period(s) during which the desired gate operates on the system. During these two types of evolution, unwanted environmental noise acts to decohere the system and increase the gate error rate. While process tomography may provide the complete map of the action of a gate operation, it does not characterize the noise sources which cause fidelity degradation and which, in principle, may differ under free- and driven-evolution conditions. In contrast, the noise sensitivity of the quantum device itself, in conjunction with tailored pulse sequences, may be used to identify spectral features of the noise and thereby elucidate the underlying noise mechanisms [2, 3].

Dynamical decoupling protocols, e.g., spin echo and its multi-pulse extensions, have been applied to mitigate dephasing during free evolution in numerous qubit modalities, including atomic ensembles [4, 5] and single atoms [6], spin ensembles [7], diamond nitrogen–vacancy centres [8, 9], semiconductor quantum dots [10, 11], and superconducting qubits [12]. The spectral filtering properties [13] of these and related sequences have proven a useful tool for performing free-evolution noise spectroscopy with superconducting qubits over a wide range of frequencies spanning millihertz [14, 15] to over 20 MHz [12]. Several groups have also made progress characterizing [12, 16, 17] and mitigating [18] decoherence during driven evolution. This case is more difficult in practice, however, because driven-evolution decoherence may be highly correlated with the operational errors that occur while implementing a particular protocol. Such correlations pose challenges for fault-tolerant quantum information processing, and it motivates the importance of characterizing the noise processes which are manifest while the qubit is under external drive.

Noise spectroscopy during driven evolution has been studied most extensively via Rabi nutation of the qubit between its ground and excited states, achieved by continuously driving the qubit with an oscillating field. Noise power at the Rabi frequency is inferred from the Rabi-oscillation-decay envelope, and the noise spectrum is reconstructed over the achievable range of Rabi frequencies by varying the field amplitude [12, 16, 17] (and F. Yoshihara et al., manuscript in preparation). There are several drawbacks to this approach, however, including the Rabi experiment's associated transverse decay, its sensitivity to low-frequency fluctuations of the Rabi frequency, the interpretation of the resultant non-exponential decay law, and the practical consideration of sampling sufficiently many points to resolve and fit accurately a decaying sinusoid oscillating at the Rabi frequency. This type of Rabi-based noise spectroscopy can, however, be generalized by preparing the qubit in an arbitrary initial state. One unique case, originating within the NMR community, is the so-called $T_{1\rho}$ experiment [16, 19–22], which measures the driven-evolution analog to $T_1$ relaxation. In $T_{1\rho}$, the driving field is applied along an axis collinear with the qubit's state-vector in the rotating frame, resulting in a condition called "spin locking." The main advantages of $T_{1\rho}$-based noise spectroscopy are that the qubit decays longitudinally from the spin-locked state with a straightforward exponential decay law (no oscillations); the decay is dominated by noise at the spin-locking frequency and, like its $T_1$ analog, is relatively insensitive to broadband low-frequency noise; and it can therefore have higher accuracy over a wider frequency range than the standard Rabi-based approach.

In this work, we develop a modified $T_{1\rho}$ pulse sequence to perform noise spectroscopy during the driven evolution of a superconducting flux qubit. The modified pulse sequence addresses several non-idealities due primarily to 1/f-type noise that may pollute the spin-locking dynamics when using the conventional $T_{1\rho}$ sequence. Using this improved sequence, we measure both the flux and tunnel-coupling noise spectra during driven evolution over the frequency range $0.1 - 200\,\mathrm{MHz}$. The observed flux-noise power spectral density (PSD) generally agrees with the $1/f$ dependence found independently using free-evolution noise spectroscopy [12]. Notably, between the experiments in Ref. [12] and those presented here, we thermal-cycled the device from $12\,\mathrm{mK}$ to $4\,\mathrm{K}$ and back again. With the higher accuracy afforded by the $T_{1\rho}$ measurement, we resolved two additional "bump"-like features in the spectrum, possibly due to a set of two-level systems (TLSs), e.g., electron spins, that were apparently activated by the thermal cycling. We further demonstrate that the underlying noise mechanisms associated with these particular spectral features are also active during free evolution by observing their signature in a time-domain echo experiment.

**Results**

**Device.** Our test device is a superconducting flux qubit (Fig. 1a), an aluminum loop interrupted by four Al-AlO$_x$-Al Josephson junctions. When an external magnetic flux $\Phi \approx \Phi_0/2$ threads the loop, the qubit's potential energy assumes a double-well profile. The wells are associated with clockwise and counterclockwise circulating currents of magnitude $I_\text{p} \approx 180\,\text{nA}$ and energies $\pm h\varepsilon/2 = \pm I_\text{p}(\Phi - \Phi_0/2)$ tunable by the applied flux. These diabatic circulating-current states tunnel-couple with a strength $\Delta = 5.4\,\text{GHz}$, and the resulting hybridized qubit states are manipulated using microwave pulses that couple to the qubit via an on-chip antenna. The qubit is embedded in a DC SQUID, a sensitive magnetometer which serves as the qubit readout. The device and measurement set-up is detailed further in Ref. [12].

**Reference frames and Hamiltonians.** In the laboratory frame and within a two-level system approximation, the qubit Hamiltonian is:

$$\hat{\mathcal{H}} = (h/2)\big[(\Delta + \delta\Delta(t))\hat{\sigma}_z + (\varepsilon + \delta\varepsilon(t))\hat{\sigma}_x + A_\text{rf}(t)\cos(2\pi\nu_\text{rf}t + \phi)\hat{\sigma}_x\big]\,, \qquad (1)$$

where $\hat{\sigma}_z$ and $\hat{\sigma}_x$ are Pauli operators. $\delta\Delta(t)$ and $\delta\varepsilon(t)$ represent random fluctuations in the effective tunnel-coupling and flux terms, respectively, and the oscillating term represents the harmonic driving field which is treated semiclassically with amplitude $A_\text{rf}(t)$, frequency $\nu_\text{rf}$, and phase $\phi$. Although noise affecting the qubit physically manifests itself in the lab frame (Eq. 1), the longitudinal ($T_1$ and $T_{1\rho}$) and transverse ($T_2$ and $T_{2\rho}$) relaxation times for free- and driven evolution may be analogously defined in the qubit-eigenbasis and rotating frames, respectively, as described below and presented in Table 1 (also see Supplementary Notes for detailed derivation and description). Their measurement may then be related back to the physical noise parameters in the laboratory frame.

Decoherence during free evolution is described by transforming Eq. 1 to the qubit's energy eigenbasis:

$$\begin{aligned}\hat{\mathcal{H}}' = (h/2)\big[&\nu_\text{q}\hat{\sigma}_{z'} + A_\text{rf}(t)\cos\theta\cos(2\pi\nu_\text{rf}t + \phi)\hat{\sigma}_{x'} \\
&+ A_\text{rf}(t)\sin\theta\cos(2\pi\nu_\text{rf}t + \phi)\hat{\sigma}_{z'} \\
&+ (\delta\Delta(t)\cos\theta + \delta\varepsilon(t)\sin\theta)\,\hat{\sigma}_{z'} + (-\delta\Delta(t)\sin\theta + \delta\varepsilon(t)\cos\theta)\,\hat{\sigma}_{x'}\big]\,,\end{aligned} \qquad (2)$$

where $\theta = \arctan(\varepsilon/\Delta)$ is the rotation angle of the quantization axis from the lab frame, and $\nu_\text{q} = \sqrt{\varepsilon^2 + \Delta^2}$ is the qubit's frequency splitting (Fig. 1b). The standard longitudinal ($T_1 \equiv 1/\Gamma_1$) and transverse ($T_2 \equiv 1/\Gamma_2$) relaxation times are defined in this frame. In our device, $T_1 \approx 12\,\mu\text{s}$ with only a weak dependence on $\varepsilon$, whereas $T_2 \approx 3\,\mu\text{s}$ at $\varepsilon = 0$ and decreases away from this point [12].

Decoherence during driven evolution (i.e., during a constant-amplitude pulse, $A_{\text{rf}}(t) = A_{\text{rf}}$) is described by a second transformation, taking Eq. 2 to a reference frame rotating at the drive frequency $\nu_{\text{rf}}$ (omitting stochastic terms):

$$\tilde{\tilde{\mathcal{H}}} = (h/2)\big[\Delta\nu\,\hat{\sigma}_Z + \nu_{\text{R}}(\cos\phi\,\hat{\sigma}_X + \sin\phi\,\hat{\sigma}_Y)\big], \tag{3}$$

where $\Delta\nu = \nu_{\text{q}} - \nu_{\text{rf}}$ is the frequency detuning between the qubit and the driving field, and $\nu_{\text{R}} = \tfrac{1}{2}A_{\text{rf}}\cos\theta$ is the Rabi frequency under resonant driving ($\Delta\nu = 0$). Here, the rotating wave approximation has been invoked by omitting the fast oscillating terms (see Supplementary Notes for a generalization).

**Analogy between free and driven evolution.** For resonant driving along the $X$-axis ($\Delta\nu = 0$, $\phi = 0$), the Hamiltonian in Eq. 3 describes a pseudospin quantized along $X$ with a level splitting equivalent to the Rabi frequency $\nu_{\text{R}}$. The qubit's Bloch vector precesses around the $X$-axis at a rate $\nu_{\text{R}}$, and its subtended angle depends on the initial state. That is, the driven evolution of the qubit state may be interpreted as the "free evolution" of a pseudospin around a static field $\nu_{\text{R}}$ (Fig. 1c). By analogy with the conventional free-evolution $T_1$ and $T_2$ times, we define longitudinal and transverse depolarization times $T_{1\rho}$ and $T_{2\rho}$ for the pseudospin with respect to this new quantization axis (see Table 1). Just as $T_1$ depends on the noise at the qubit frequency $\nu_{\text{q}}$ that is transverse to the $z'$-axis (Eq. 2), so is $T_{1\rho}$ determined by the noise at the pseudospin's level splitting $\nu_{\text{R}}$ that is transverse to the $X$-axis (Eq. 3). For completeness, the measurement and value of the rotating-frame decoherence time $T_{2\rho}$ ("free-induction" time in the rotating frame) is equivalent to the conventional Rabi experiment and associated decay time $T_{\text{Rabi}}$ (Supplementary Fig. S2b). More detailed discussion about the analogy between free evolution and driven evolution in the rotating frame can be found in Supplementary Notes.

The connection between the driven-evolution times ($T_{1\rho}$, $T_{2\rho}$), the free-evolution times ($T_1$, $T_2$), and the PSD of the physical noise terms ($\delta\Delta(t)$, $\delta\varepsilon(t)$) is formally made using the generalized Bloch equations (GBE) [23, 24] and the transformations between Eqs. 1-3. The exponential relaxation rate during weak and resonant driving is

$$\Gamma_{1\rho} = \frac{1}{T_{1\rho}} = \frac{1}{2}\Gamma_1 + \Gamma_\nu, \tag{4}$$

where

$$\Gamma_\nu = \frac{1}{2} S_{z'}(\nu_{\rm R}) = \frac{1}{2}\big[\cos^2\theta\, S_\Delta(\nu_{\rm R}) + \sin^2\theta\, S_\varepsilon(\nu_{\rm R})\big] \ . \qquad (5)$$

$S_\lambda(f) = (2\pi)^2 \int_{-\infty}^{\infty} dt \frac{1}{2}\langle\lambda(0)\lambda(t) + \lambda(t)\lambda(0)\rangle \exp(-i2\pi f t)$ stands for the symmetrized PSD of $\lambda$-noise, and $\lambda \in \{\Delta, \varepsilon \ldots \text{ or } x', y', z' \ldots\}$ indicates either the fluctuating part of the system Hamiltonian or the axis along which those fluctuations occur. The relaxation rate $\Gamma_{1\rho}$ arises from two terms related to noise in the qubit eigenbasis (Eq. 2). The first term ($\Gamma_1$ term) is associated with transverse noise along the $x'$-axis at frequency $\nu_{\rm q} \pm \nu_{\rm R}$ (which appears as $\nu_{\rm R}$ in the rotating frame). The second term ($\Gamma_\nu$ term) is the longitudinal noise along the $z'$-axis at the drive frequency $\nu_{\rm R}$, which is directly transferred into the rotating frame. This latter term provides a direct method to extract the noise PSD at the locking Rabi frequency by measuring relaxation during driven evolution via a $T_{1\rho}$ experiment.

**Advantages of $T_{1\rho}$ over Rabi spectroscopy.** The $T_{1\rho}$ experiment, when utilized as a noise spectrum analyzer to extract the PSD, has inherent advantages over direct Rabi-based techniques [12, 17] (and F. Yoshihara et al., in preparation). One major shortcoming of the Rabi method is its vulnerability to low-frequency fluctuation $\delta\nu_{\rm R}$ of the effective Rabi frequency. First, the Rabi frequency is first-order sensitive to fluctuations of the driving-field amplitude $\delta\nu_{\rm R}$ (proportional to $\delta A_{\rm rf}$), whose RMS fluctuation is found in this case proportional to $\nu_{\rm R}$ ($\delta\nu_{\rm R} \approx 0.06\%\, \nu_{\rm R}$) [18]. Hence, the temporal $\nu_{\rm R}$ inhomogeneity becomes important when the driving is strong (large $\nu_{\rm R}$). In addition, low-frequency fluctuation of the qubit's level splitting ($\delta\nu$) modulates the effective Rabi frequency to second order, adding to the inhomogeneous broadening [12]. The effect dominates the Rabi decay when the driving is weak enough (small $\nu_{\rm R}$, as the second-order contribution $\langle\delta\nu^2\rangle/\nu_{\rm R}$ exceeds decay rates from other sources) (Table 1). These two effects are analogous to the scenarios of free-evolution dephasing due to low-frequency noise via linear and quadratic coupling, respectively [16]. In contrast, the locking dynamics in the $T_{1\rho}$ experiment largely reduce its sensitivity to these fluctuations, and thus substantially improves the accuracy of extracted noise PSD, especially at both high and low frequency regimes.

In addition, $T_{1\rho}$ has a simple exponential decay law, whereas the Rabi decay law can be complicated by its sinusoidal nutation in conjunction with a non-exponential decay envelope related to its sensitivity to low-frequency noise (see Table 1), both of which pose a challenge to high-precision fitting. A $T_{1\rho}$ measurement requires far fewer datapoints than would a Rabi decay envelope on a sinusoid oscillating at the Rabi frequency. These properties make the $T_{1\rho}$ method more accurate and experimentally efficient.

**Conventional $T_{1\rho}$ sequence (SL-3).** The measurement of the rotating-frame relaxation time $T_{1\rho}$ begins by preparing the qubit state to be either parallel or anti-parallel with $X$ (collinear with the driving field), corresponding to the pseudospin's ground and excited state, respectively. This results in the so-called "spin-locking" condition, a reference to the pseudospin remaining aligned in the absence of noise with the static driving field $\nu_{\text{R}}$ in Eq. 3 (see Fig. 1e and Supplementary Fig. S2a). The measurement of the relaxation time from the spin-locked condition to a depolarized steady-state is the so-called $T_{1\rho}$ experiment.

The original $T_{1\rho}$ sequence as developed in NMR and applied to superconducting qubits was a three-pulse sequence [25], two $\pi/2$ pulses separated by a 90°-phase-shifted continuous driving pulse (labelled here "SL-3", see Fig. 1d). The phase shift is implemented by altering the driving-field phase $\phi$ via standard IQ mixing techniques. The dynamics can be visualized with the assistance of the Bloch sphere within the rotating frame (Fig. 1e). For convenience, we use $\{X, Y, Z\}$ to denote rotating frame coordinates, pulse phase, and qubit states. The first $\pi/2_{\overline{Y}}$ pulse (bar indicates negative or 180-degrees-rotated orientation) brings the qubit from its ground state to the equator, parallel with $X$ (step I). The driving field is then applied along X and thereby aligned with the qubit state, so that the qubit is effectively locked in this orientation and experiences relaxation $\Gamma_{1\rho}$ only (steps II and III). After a finite duration $\tau$, the remaining polarization along $X$ is projected, by the last $\pi/2_{\overline{Y}}$ pulse, back along $Z$ for readout (step IV). The long-time, steady-state population carries no net $X$-polarization, because the device is operated in the limit $\nu_{\text{R}} \ll k_{\text{B}}T/h = 1.3\,\text{GHz}$, where $T \approx 65\,\text{mK}$ is the effective device temperature obtained by measuring the SQUID's switching-current distribution. The $\overline{Y}$-$X$-$\overline{Y}$ phase order applied here and the $X$-$Y$-$X$ one in [25] are effectively the same, since only the relative phase is important. We repeat the same sequence on a nominally identically prepared system a few thousand times to improve state estimation. Hence, decoherence of this system should be understood in the sense of a temporal ensemble average.

**Modified $T_{1\rho}$ sequence (interleaved SL-5a and SL-5b).** In solid-state systems, the recorded $T_{1\rho}$ decay signal using the original (SL-3) sequence may suffer from several types of system imperfection, in particular, signal distortion due to low-frequency noise via other channels. One example of a measured $T_{1\rho}$ trace is shown in Fig. 2b (brown trace). Another example can be found in Ref. [25], an early demonstration of spin-locking implemented on a quantronium qubit. A prominent feature in these examples is the presence of unwanted Rabi-frequency oscillations on the back of an otherwise exponential decay function. We excluded the possibility of a constant frequency detuning in our experiment, which could effectively separate the driving field and qubit state during the locking period and lead to oscillations. Rather, we determined that the oscillations were due to dephasing during the intervals between adjacent pulses. In our experiments, we use Gaussian pulses with $\sim 10\,\mathrm{ns}$ width for the $\pi/2$ pulses and Gaussian rise-and-fall for the long driving pulse. To clearly separate pulses, there is a practically unavoidable duration of free evolution between pulses (Fig. 1d). Depending on the length $\tau$ of the second pulse, phase diffusion accumulated during the first interval will sometimes be doubled (for $\nu_{\mathrm{R}}\tau = 1, 2, 3\ldots$), and sometimes be refocused (for $\nu_{\mathrm{R}}\tau = \frac{1}{2}, 1\frac{1}{2}, 2\frac{1}{2}\ldots$), producing those oscillations. In the presence of this noise, the locking protocol is compromised to a differing degree for the temporal ensemble elements, making the observed decay a complicated combination of precession, relaxation and decoherence processes. To correct for it, we developed a modified five-pulse sequence (labelled "SL-5a," see Fig. 2a), which adds one $\pi_X$ pulse in the middle of each interval in SL-3. The added pulses refocus the phase diffusion as is done in a spin-echo sequence [26]. Relaxation obtained using SL-5a is shown in Fig. 2b, and it shows a clear improvement (no oscillation) over SL-3.

In several measured traces, we observe a second type of signal distortion—random fluctuations at the few-percent level on the exponential signal, and at a time scale of minutes or tens of minutes—e.g., Fig. 2c (blue trace). As we describe in detail below, low-frequency fluctuations of the frequency detuning result in the "shivering" of the effective Rabi field and, thereby, fluctuations in the observed decay signal. We first describe how we corrected the distortion, and then explain its origin. The solution is to use a complementary pulse sequence (labelled "SL-5b," see Fig. 2a) together with SL-5a. The new sequence is nominally identical to SL-5a, except that we replace the two $\pi/2_{\overline{Y}}$ pulses with two $\pi/2_Y$ pulses, such that the qubit Bloch vector is anti-parallel with the driving axis for SL-5b. Ideally, both sequences should give the same decay function. However, when we apply the sequences in an interleaved sampling order (inner loop: alternate between SL-5a and SL-5b; outer loop: step $\tau$), the anti-symmetry of the signal fluctuation, slow compared with the collection time at each value $\tau$, is captured by the twin sequences, as shown by the example in Fig. 2c. We then recover the smooth exponential decay by taking the average of the two traces.

Our ability to recover a smooth exponential decay is closely related to the ultra-slow (∼minutes) fluctuation of frequency detuning $\delta\nu$. First, the occurrence of the signal distortion at large $\tau$ indicates a deviation of the steady-state population from zero $X$-polarization. Assuming small detuning ($\Delta\nu \ll \nu_R$), to first order, the steady-state $X$-polarization [24] (and Supplementary Notes), in the practical limit $\nu_R \ll k_B T/h \ll \nu_q$, is

$$\langle \hat{\sigma}_X \rangle^{\mathrm{ss}} = -\frac{\sin\eta\, S_{x'}(\nu_q)}{\frac{1}{2} S_{x'}(\nu_q) + S_{z'}(\nu_R)} \;, \qquad (6)$$

where $\eta = \arctan(\Delta\nu/\nu_R)$ is the angle of the driving field with respect to X due to frequency detuning (see Fig. 2d). Equation 6 is applicable to both SL-5a and SL-5b, but the difference in the last back-projection pulse gives rise to a differential signal response to the non-zero $X$-polarization. Therefore, fluctuation of the frequency detuning, to first order, causes fluctuation of the steady-state longitudinal polarization and the anti-symmetric feature of the traces in Fig. 2c. The relaxation dynamics in both sequences with finite detuning are sketched in the right panel of Fig. 2d. Ignoring the process at small $\tau$ where the much faster $T_{2\rho}$ process is involved, the normalized decay signal, to first order of $\eta$, can be approximated as $P_a(\tau) = \frac{1}{2}(1 - \langle\hat{\sigma}_X\rangle^{\mathrm{ss}})(1 - e^{-\Gamma_{1\rho}\tau})$ for SL-5a, and $P_b(\tau) = \frac{1}{2}(1 + \langle\hat{\sigma}_X\rangle^{\mathrm{ss}})(1 - e^{-\Gamma_{1\rho}\tau})$ for SL-5b. The average $P(\tau) = \frac{1}{2}(P_a(\tau) + P_b(\tau)) = \frac{1}{2}(1 - e^{-\Gamma_{1\rho}\tau})$ becomes independent of detuning, and recovers the underlying relaxation signal.

**Noise spectroscopy during driven evolution.** We use the modified $T_{1\rho}$ pulse sequence to implement noise spectroscopy during driven evolution (Fig. 3). To derive the $\Delta$-noise PSD, we perform a driven-evolution $T_{1\rho}$ measurement for various $\nu_R$ at $\varepsilon = 0$, where the qubit is first-order insensitive to $\varepsilon$ noise, as well as the standard (free-evolution) inversion-recovery $T_1$-measurement [26]. Both $T_1$ and $T_{1\rho}$ traces are fit to extract the exponential decay rates $\Gamma_1$ and $\Gamma_{1\rho}$, which give the $\Delta$-noise PSD by $S_\Delta(\nu_R) = S_{z'}(\nu_R) = 2\Gamma_{1\rho} - \Gamma_1$. For $\varepsilon$-noise PSD, we perform the same experiments at $\varepsilon \neq 0$, where the qubit becomes predominantly sensitive to $\varepsilon$ noise. We then have $S_\varepsilon(\nu_R) = [S_{z'}(\nu_R) - \cos^2\theta S_\Delta(\nu_R)]/\sin^2\theta = [2\Gamma_{1\rho} - \Gamma_1 - \cos^2\theta S_\Delta(\nu_R)]/\sin^2\theta$, where $S_\Delta(\nu_R)$ is previously measured. The locking Rabi frequency, $\nu_R$, is determined by measuring Rabi oscillations with the same driving-field amplitude. The lower limit ($\sim 100\,\mathrm{kHz}$) of the spectrum is determined by the $\Delta\nu$ inhomogeneity $\langle\delta\nu^2\rangle$, which breaks the locking condition when driving is too weak. On the other hand, an undetermined effect (presumably heating due to strong driving) which distorts the SQUID's switching-current distribution prevents us from probing higher frequency than $\sim 200\,\text{MHz}$. We believe the limitation is unrelated to the SQUID plasma mode, which is around $\sim 2.3\,\text{GHz}$ for this device.

In this device, the measured $\Delta$-noise spectrum is generally white. Since the extrapolated $1/f$ $\Delta$-noise from low-frequency measurements crosses the white noise level around the lower limit, we are unable to determine whether the $1/f$ noise extends to this regime or has a cut-off [14]. The noise could be due to Johnson noise from charge or critical current fluctuations, or due to thermal photon noise [27] originating from the $LC$ resonator associated with the readout SQUID which inductively couples to the qubit (Fig. 1a).

The $\varepsilon$-noise spectrum exhibits a $1/f^\alpha$ dependence with $\alpha = 0.9$, a value that generally agrees with independently measured free-evolution noise spectroscopy using a Ramsey $(1\,\text{mHz} - 100\,\text{Hz})$ [14] and CPMG $(0.1 - 20\,\text{MHz})$ approach [12]. Although not necessarily a universal result, this suggests that driven evolution does not activate $\varepsilon$-noise sources for this device. In general, the $T_{1\rho}$ spectroscopy confirms these results with better accuracy and extends the frequency range by 5x (presumably limited in this device by a heating effect at high driving fields). Note that measuring at higher frequency $(0.1 - 1\,\text{GHz})$ is feasible in principle by using this technique.

**Spectral and temporal signatures of coherent TLSs.** In contrast to previous free-evolution measurements, we observe on the $1/f$ trend two clear "bump-like" features, one around $1\,\text{MHz}$ and the other around $20\,\text{MHz}$. These features appeared after thermal-cycling the device from base temperature to $4.2\,\text{K}$ and back again, which likely activated the fluctuators. The one at $1\,\text{MHz}$ has a clear peak above the general $1/f$ background. This excludes the possibility of random telegraph noise (RTN), possibly generated by some microscopic two-level fluctuators (TLFs), as the cause of this bump, since RTN would produce a Lorentzian spectrum centered at zero frequency [28, 29]. To fit the main features, we assume that the $\varepsilon$-noise PSD is a combination of the $1/f^{0.9}$ line and two Lorentzians. Both bumps are better fit to Lorentzians centered around $1\,\text{MHz}$ and $20\,\text{MHz}$ than those centered at zero frequency. One example is shown by the black line in Fig. 3.

We find a signature of the $1\,\text{MHz}$ bump feature in a free-evolution, time-domain echo experiment. Fig. 4 shows the spin-echo phase-decay as measured at several flux-bias points where the qubit coherence is mostly limited by $\varepsilon$ noise. Under a Gaussian-noise assumption, the echo sequence is most sensitive to the noise whose frequency is about the inverse of the total free-evolution time of the sequence. The echo decay functions all exhibit a clear "dip" feature around $1\,\mu$s, corresponding to the spectral feature around $1\,$MHz. To model this result, we use the previously assumed PSD function $S'(f)$ ($1/f^{0.9}$ plus two Lorentzians) with tunable parameters to reproduce the echo-decay signal by the filter function method [12, 16, 30]. The agreement is good, and bump parameters estimated from the temporal echo measurement are within 30% of a direct fit to the driven-evolution spectral data. In general, we are able to predict the phase decay from the $T_{1\rho}$ spectroscopy results, and, in turn, justify the performance of the $T_{1\rho}$ spectroscopy method on a superconducting qubit. In the case of the 20-MHz Lorentzian, simulations indicate that the echo signature corresponding to this broad, relatively small feature in the noise spectrum would yield only a small deviation from the spin-echo phase-decay function with no discernable "dip." The temporal resolution in Fig. 4 is furthermore insufficient to resolve such a deviation. We note that on a subsequent cooldown, with the qubit exhibiting a similarly small spectral feature, we did observe a small distortion (but no "dip") in the expected spin-echo phase-decay function.

It is possible that the observed Lorentzian bumps are generated by two sets of coherent TLSs [31], such as electron spins. In a magnetic environment, each spin executes Larmor precession at a frequency proportional to its local field. Its transverse component couples to the qubit's flux degree of freedom with a strength depending on geometric relations among the device, the field and the spin [32], and adds to a net fluctuating flux whose PSD reflects the spectral (Larmor-frequency) distribution of spins, even though the ensemble has already lost its coherence. The width of the Lorentzian indicates in part the field inhomogeneity. Assuming uncorrelated surface electron spins and an average coupling strength of $2.7 \times 10^{-8}\Phi_0 \cdot \mu_{\rm B}^{-1}$ from simulation, both Lorentzian bumps correspond to $\sim 10^6$ spins. The number scales down in the presence of a high degree of spin order [33]. Moreover, if we assume the spins feel the same magnetic field as the qubit, about $0.5\,$mT ($B_0 = \Phi/A_{\rm L}$, where $\Phi = \Phi_0/2$ is the flux through the qubit loop and $A_{\rm L} \approx 2\,\mu$m$^2$ is the loop size), the corresponding electron-spin Larmor frequency is $14\,$MHz. This is only a rough estimate, as screening due to superconducting metal may lead to a large variation of the field at various locations in the vicinity of the metal surface, and other locations for these spins, e.g., superconductor insulator boundary [34] and substrate, are also possible. Nonetheless, the crude estimate is consistent with our observations.

**Discussion**

The modified $T_{1\rho}$ method presented here, in conjunction with existing free-evolution spectroscopy methods, enables the characterization of the noise sources relevant for both driven- and free-evolution quantum control methods. Although many noise sources may be manifest similarly during both driving conditions, in principle, the noise in these two cases need not be identical. For example, driven evolution may activate certain noise sources that are otherwise dormant during free-evolution. In this demonstration with a flux qubit, the flux and tunnel-coupling noise spectra reveal more detailed information and cover a wider frequency range, a 10x increase in this device, as compared with our previous free-evolution (CPMG spectroscopy) and driven evolution (Rabi spectroscopy) methods [12]. We observed the spectral features of two sets of coherent TLSs in the environment, possibly due to effective electron spins on the metal surface, which are active during driven evolution. We could furthermore observe a temporal signature of one of these fluctuators in a free-evolution echo experiment, indicating that it is active during both driven- and free evolution. This type of noise characterization serves as an important step towards the engineered mitigation of decoherence through improved materials, fabrication, control sequences, and qubit design. Our modifications to the $T_{1\rho}$ pulse sequence target the unwanted effects of low-frequency noise, opening this method to the multiple qubit modalities that are subject to such noise.

**Methods**

**Description of the qubit.** We fabricated our device at NEC, using the standard Dolan angle-evaporation deposition process of Al–AlO$_x$–Al on a SiO$_2$/Si wafer.

Our persistent-current qubit [12, 35] consists of a superconducting loop with diameter $d \sim 2\,\mu$m, interrupted by four Josephson junctions. Three of the junctions each have the Josephson energy $E_{\rm J} = h \times 210\,{\rm GHz}$, and charging energy $E_{\rm C} = h \times 4\,{\rm GHz}$; the fourth is smaller by a factor $\alpha = 0.54$.

The qubit is embedded in a hysteretic DC SQUID, a sensitive magnetometer which is used for the qubit readout. It has a critical current $I_{\rm c,sq} = 4.5\,\mu{\rm A}$; normal resistance $R_{\rm N} = 0.25\,{\rm k}\Omega$; mutual qubit–SQUID inductance $M_{\rm Q-S} = 21\,{\rm pH}$; on-chip shunt capacitors $C_{\rm sh} \approx 10\,{\rm pF}$, inductors $L_{\rm sq} \approx 0.1\,{\rm nH}$, and bias resistors $R_I = 1\,{\rm k}\Omega$ and $R_V = 1\,{\rm k}\Omega$. The shunt capacitors bring the plasma frequency down to $f_\mathrm{p} = 2.1\,\mathrm{GHz}$, which is much higher than the frequencies of the coherent fluctuators presumably related to the frequency- and time-domain results.

**Pulse generation and calibration.** The r.f. microwave pulses are created by mixing the in-phase and quadrature pulse envelopes, generated by a Tektronix 5014 Arbitrary Waveform Generator (AWG), with a continuous wave, provided by an Agilent E8267D PSG Vector Signal Generator. We use the internal I/Q mixer of the PSG for the envelope mixing. All pulses are further gated in order to eliminate leakage.

Imperfections in the electronics and coaxial cables will cause pulse distortions, especially for short pulses (a few nanoseconds long). To ensure that the pulses we send to the cryostat are free from distortions, we first determine the transfer function $H_\mathrm{ext}$ with a high-speed oscilloscope, and then use $H_\mathrm{ext}^{-1}$ to correct for imperfections in the AWG and in the I/Q mixers [36]. This set-up allows us to create well-defined Gaussian-shaped microwave pulses with pulse widths as short as 2.5 ns.

---

**Acknowledgements**

We appreciate A. Kerman, P. Krantz, L. Levitov, S. Lloyd, and T. Yamamoto for helpful discussions and K. Harrabi for assistance with device fabrication. We thank R. Wu for support in this work. This work was sponsored in part by the US Government, the Laboratory for Physical Sciences, the U.S. Army Research Office (W911NF-12-1-0036), the National Science Foundation (PHY-1005373), the Funding Program for World-Leading Innovative R&D on Science and Technology (FIRST), NICT Commissioned Research, MEXT kakenhi 'Quantum Cybernetics', Project for Developing Innovation Systems of MEXT. Opinions, interpretations, conclusions and recommendations are those of the author(s) and are not necessarily endorsed by the U.S. Government.

**Author Contributions**

F. Yoshihara and Y.N. designed and fabricated the device. F. Yan, S.G. and J.B. performed the experiments and contributed to the software infrastructure. F. Yan analyzed the data. F. Yan and W.D.O. wrote the paper with feedback from all authors. W.D.O., D.G.C. and T.P.O. supervised the project. All authors contributed to discussions during the conception, execution, and interpretation of the experiments.

|  | **Decoherence during free evolution** | **Decoherence during driven evolution** |
|---|---|---|
| Working frame | Qubit frame | Rotating frame |
| Quantization axis | $z'$-axis | $X$-axis |
| Nutation freq. | Level splitting $\nu_q$ | Rabi frequency $\nu_R$ |
|  | Longitudinal relaxation ||
| Method | Inversion recovery [26, 37] | Spin locking - $T_{1\rho}$ ([16], this work) |
| Longitudinal decay | $z'$-axis | $X$-axis |
| Decay law | $\exp\{-\Gamma_1 \tau\}$ | $\exp\{-(\frac{1}{2}\Gamma_1 + \Gamma_\nu)\tau\}$ |
| Noise of interest | $\Gamma_1 = \frac{1}{2} S_{x'}(\nu_q)$ | $\Gamma_\nu = \frac{1}{2} S_{z'}(\nu_R)$ |
| Noise susceptibility | $S_{x'}(\nu_q)$ | $S_{x'}(\nu_q), S_{z'}(\nu_R)$ |
|  | Transverse decoherence ||
| Method | Ramsey [26, 35], CPMG [12] | Rabi [12], Rotary echo [18] |
| Transverse decay | $x'-y'$ plane | $Y-Z$ plane |
| Decay law | $\exp\{-\frac{1}{2}\Gamma_1 \tau - \frac{(2\pi)^2}{2}\langle \delta\nu^2\rangle \tau^2\}$ $\times \cos(2\pi\Delta\nu\tau)$ | $\exp\{-(\frac{3}{4}\Gamma_1 + \frac{1}{2}\Gamma_\nu)\tau - \frac{(2\pi)^2}{2}\langle \delta\nu_R^2\rangle \tau^2\}$ $\times [1 + (2\pi \frac{\langle \delta\nu^2\rangle}{\nu_R}\tau)^2]^{-1/4}\cos(2\pi\nu_R\tau)$ |
| Noise of interest | $\langle \delta\nu^2 \rangle$ | $\Gamma_\nu = \frac{1}{2} S_{z'}(\nu_R)$ |
| Noise susceptibility | $S_{x'}(\nu_q)$, $\Delta\nu$ inhom. | $S_{x'}(\nu_q), S_{z'}(\nu_R)$, $\nu_R$ and $\Delta\nu$ inhom. |

**Table** 1: Comparison of decohering properties of a flux qubit during free and driven evolution (weak and resonant).

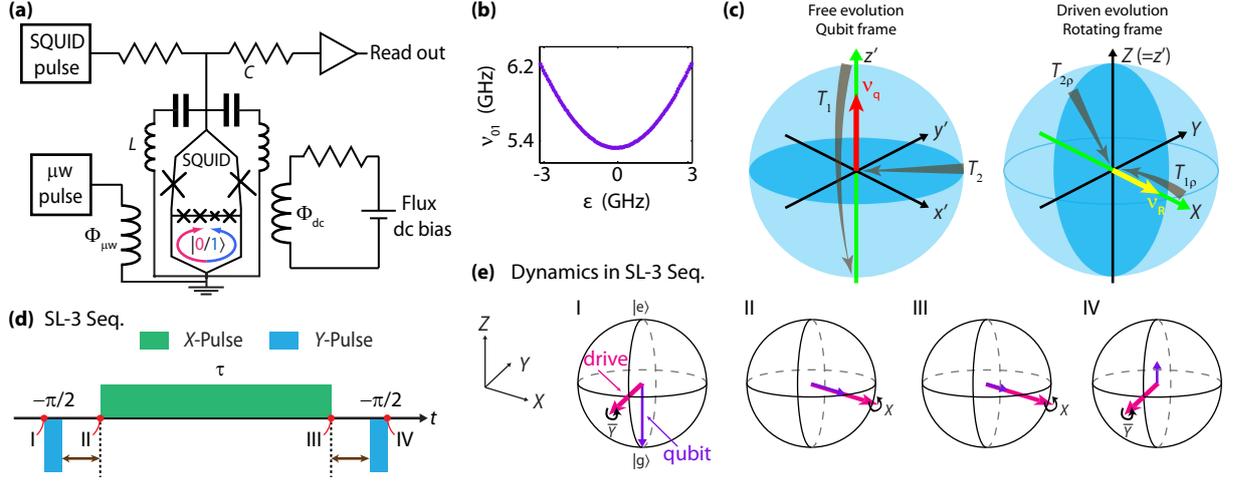

**Figure** 1: **Device schematic and spin-locking dynamics.** (a) Schematic of the flux qubit with measurement set-up. (b) Measured qubit frequency versus flux bias. (c) Analogy between free- and driven evolution. Red and yellow arrows indicate the quantizing field, respectively. The longitudinal axes are highlighted by green arrows, and the transverse planes are highlighted by the blue disks. Grey arrows indicate the longitudinal ($T_1$ and $T_{1\rho}$) and transverse ($T_2$ and $T_{2\rho}$) depolarization. (d) Standard three-pulse spin-locking sequence (SL-3). Green and blue indicate $X$- and $Y$-pulse, respectively. (e) Bloch sphere representation of the rotating-frame qubit dynamics under SL-3. The purple arrows are the qubit's state polarization, while the magenta arrows indicate the driving-field orientation. The qubit is initially prepared in its ground state (I). The first $\pi/2_{\overline{Y}}$ pulse rotates the qubit by 90° into the equatorial plane (II). The second 90°-phase-shifted continuous driving pulse of duration $\tau$, is then aligned with the qubit state, effectively locking the qubit along $X$. During the pulse, the qubit undergoes relaxation in this rotating frame towards its steady state (III). The final $\pi/2_{\overline{Y}}$ pulse projects the remaining polarization onto $Z$ ($=z'$) for readout (IV).

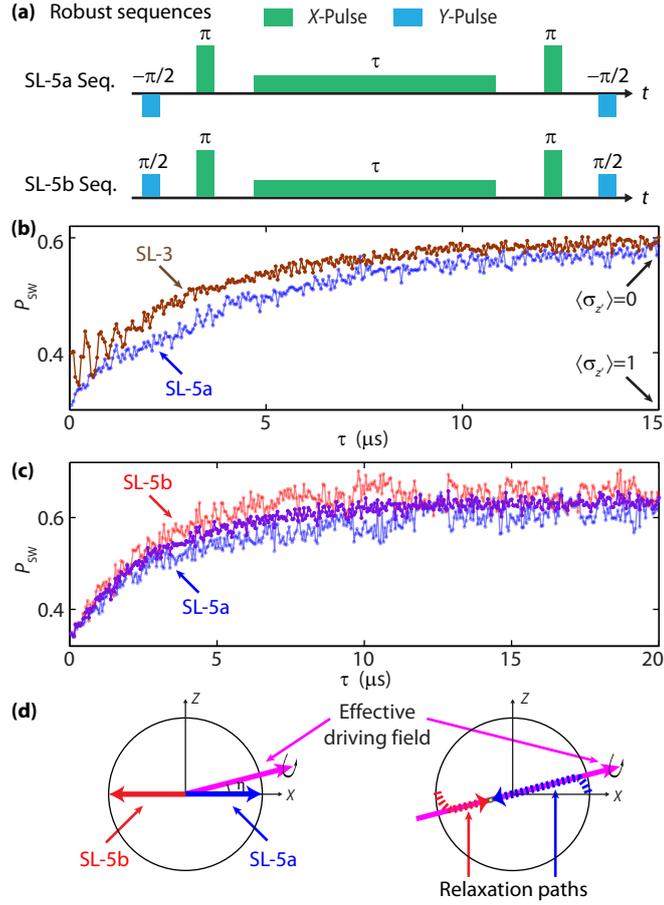

**Figure** 2: **Robust spin-locking sequences.** (a) Modified five-pulse sequences (SL-5a and SL-5b), robust in the presence of $\delta\varepsilon$ and $\delta\Delta$ low-frequency noise. (b) Comparison of measured signals by SL-3 (brown) and SL-5a (blue). Both traces are taken at $\varepsilon = 192\,\mathrm{MHz}$ and $\nu_\mathrm{R} = 3\,\mathrm{MHz}$. $P_\mathrm{SW}$ is the measured switching probability of the readout SQUID, and is linear with the qubit's state population. (c) Comparison of measured signals by SL-5a (blue), SL-5b (red) and their average (purple). Both traces are taken at $\varepsilon = 960\,\mathrm{MHz}$ and $\nu_\mathrm{R} = 24\,\mathrm{MHz}$. The data are taken in an interleaved order between blue and red traces. (d) Cross-section sketch of the qubit dynamics under the SL-5a and SL-5b protocols, with the driving field (magenta arrow) tilted by an angle $\eta$ ($\eta > 0$ in this example) due to slow fluctuations of the frequency detuning. On the left are the qubit states right before the locking pulse; on the right, their relaxation paths during the locking period are indicated by the dashed lines. The steady-state solution (the contacting point of blue and red dash lines) deviates away from $\langle \hat{\sigma}_X \rangle = 0$ when $\eta \neq 0$.

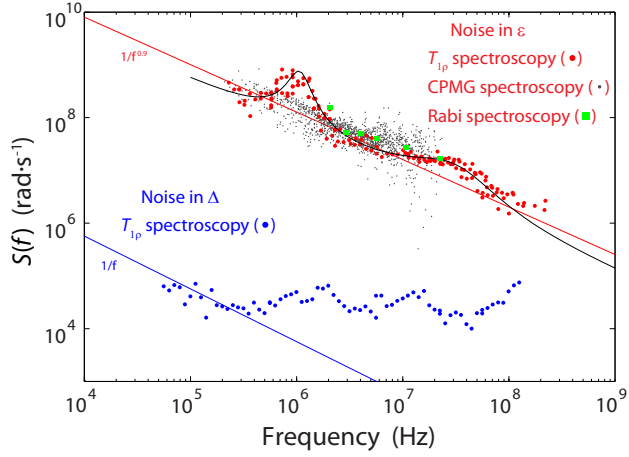

**Figure** 3: **Noise spectroscopy during driven evolution.** $\varepsilon$ (red dots) and $\Delta$ (blue dots) noise spectroscopy by the $T_{1\rho}$ method. Here, previous results on $\varepsilon$ noise by the CPMG (grey dots) and Rabi (green dots) methods measured on the same device are plotted for comparison. From the fitting results, the error rate in the worst case is about 30% for $\varepsilon$ noise, and 70% for $\Delta$ noise. Previous experiments and the $T_{1\rho}$ experiment were separated by six months and a 4K warm-up. The red and blue solid lines are the $1/f^\alpha$ power laws, extrapolated from separate low-frequency noise measurements [14]. The black line represents the function $S'(f) = A/f^{0.9} + S_1 \mathcal{L}(f; F_1, W_1) + S_2 \mathcal{L}(f; F_2, W_2)$, where $\mathcal{L}(f; F, W) = W^2/((f - F)^2 + W^2)$, and $A = (2\pi\, 0.65 \times 10^6)^2\, (\text{rad·s}^{-1})^2$, $S_1 = 7.0 \times 10^8\, \text{rad·s}^{-1}$, $F_1 = 1.05\, \text{MHz}$, $W_1 = 0.25\, \text{MHz}$, $S_2 = 1.2 \times 10^7\, \text{rad·s}^{-1}$, $F_2 = 20\, \text{MHz}$, $W_2 = 25\, \text{MHz}$.

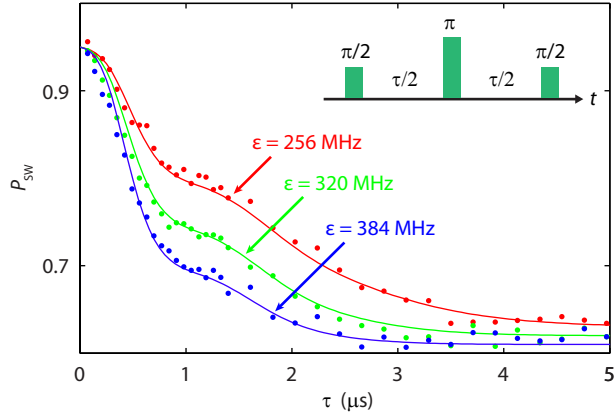

**Figure** 4: **Temporal signature of TLS during free-evolution.** Echo phase-memory decay (dots) measured at $\varepsilon = 256\,\text{MHz}$ (red), $320\,\text{MHz}$ (green) and $384\,\text{MHz}$ (blue), and corresponding simulation (solid lines) with assumed flux noise PSD function, $S'(f) = A/f^{0.9} + S_1 \mathcal{L}(f; F_1, W_1)$, where $A = (2\pi\, 0.9 \times 10^6)^2\,(\text{rad}\cdot\text{s}^{-1})^2$, $S_1 = 9.0 \times 10^8\,\text{rad}\cdot\text{s}^{-1}$, $F_1 = 1.25\,\text{MHz}$, $W_1 = 0.35\,\text{MHz}$. The temporal signature of the Lorentzian at $1\,\text{MHz}$ is essentially independent of the $\varepsilon$ bias. Inset is the echo pulse sequence. Here, $\tau$ is the total free evolution time in the sequence. Note that the $20\,\text{MHz}$ bump has little influence at these bias points.

# Supplementary information: Rotating-frame relaxation as a noise spectrum analyzer of a superconducting qubit undergoing driven evolution

Fei Yan[1*], Simon Gustavsson[2], Jonas Bylander[2], Xiaoyue Jin[2], Fumiki Yoshihara[3], David G. Cory[1,4,5], Yasunobu Nakamura[3,6], Terry P. Orlando[2], and William D. Oliver[2,7]

[1]*Department of Nuclear Science and Engineering,*
*Massachusetts Institute of Technology (MIT), Cambridge, MA 02139, USA*
[2]*Research Laboratory of Electronics, MIT, Cambridge, MA 02139, USA*
[3]*The Institute of Physical and Chemical Research (RIKEN), Wako, Saitama 351-0198, Japan*
[4]*Institute for Quantum Computing and Department of Chemistry,*
*University of Waterloo, Waterloo, Ontario N2L 3G1, Canada*
[5]*Perimeter Institute for Theoretical Physics, Waterloo, ON, N2J 2W9, Canada*
[6]*Research Center for Advanced Science and Technology,*
*The University of Tokyo, Komaba, Meguro-ku, Tokyo 153-8904, Japan*
[7]*MIT Lincoln Laboratory, 244 Wood Street, Lexington, MA 02420, USA*

———

* fyan@mit.edu

**(a)** Laboratory frame

**(b)** Qubit frame

**(c)** Rotating frame ($\varphi=0$)

**(d)** Rotating frame (on-resonance)

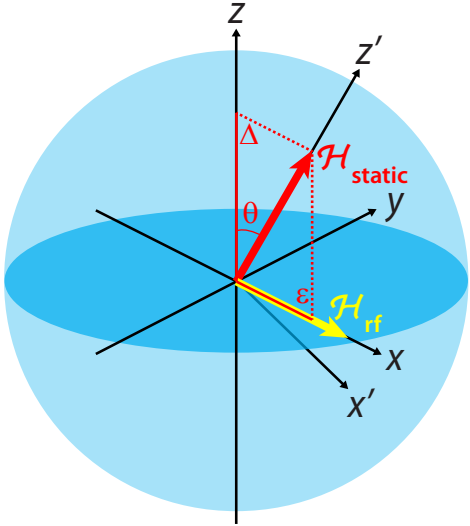
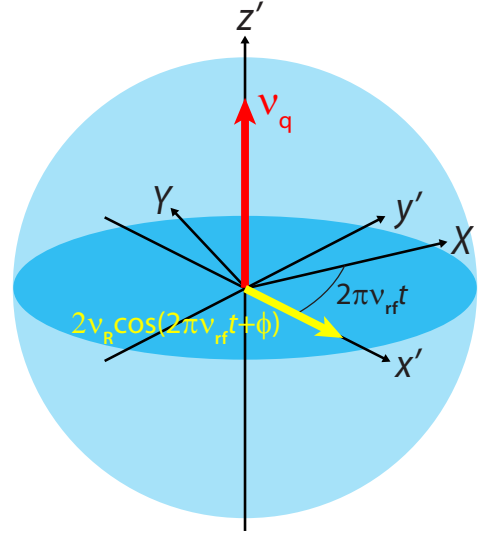
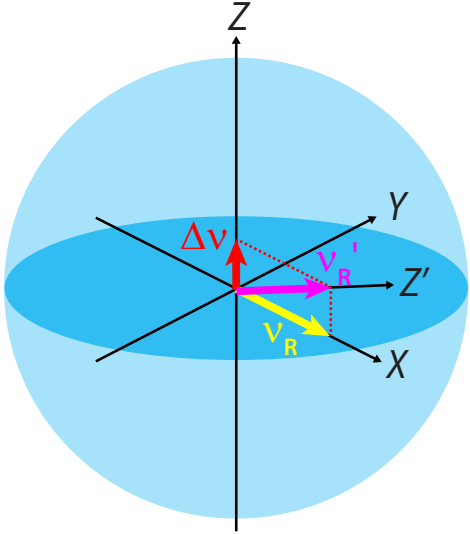
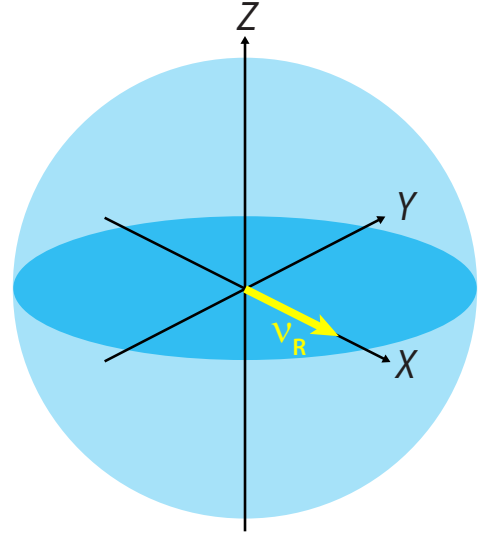

**Supplementary Figure S1. System Hamiltonian for different reference frames.** **(a)** Lab frame. $\hat{\mathcal{H}}_{\text{static}} = \Delta \hat{\sigma}_z + \varepsilon \hat{\sigma}_x$, $\hat{\mathcal{H}}_{\text{rf}} = A_{\text{rf}} \cos(2\pi\nu_{\text{rf}}t + \phi)\hat{\sigma}_x$, and $\theta$ indicates the tilting of quantization axis when transformed into the qubit frame. **(b)** Qubit frame. The red arrow ($\nu_{\text{q}}$) is the free-evolution quantizing field, and $2\pi\nu_{\text{rf}}t$ indicates the time-dependent transformation into the rotating frame. **(c)** Rotating frame ($\phi = 0$). The pink arrow ($\nu'_{\text{R}}$) indicates the effective driven-evolution quantizing (Rabi) field in the rotating frame in the presence of a finite detuning $\Delta\nu$. **(d)** Rotating frame (same as (c) but $\Delta\nu = 0$). The yellow arrow ($\nu_{\text{R}}$) indicates the resonantly driven-evolution quantizing field, or the pseudospin.

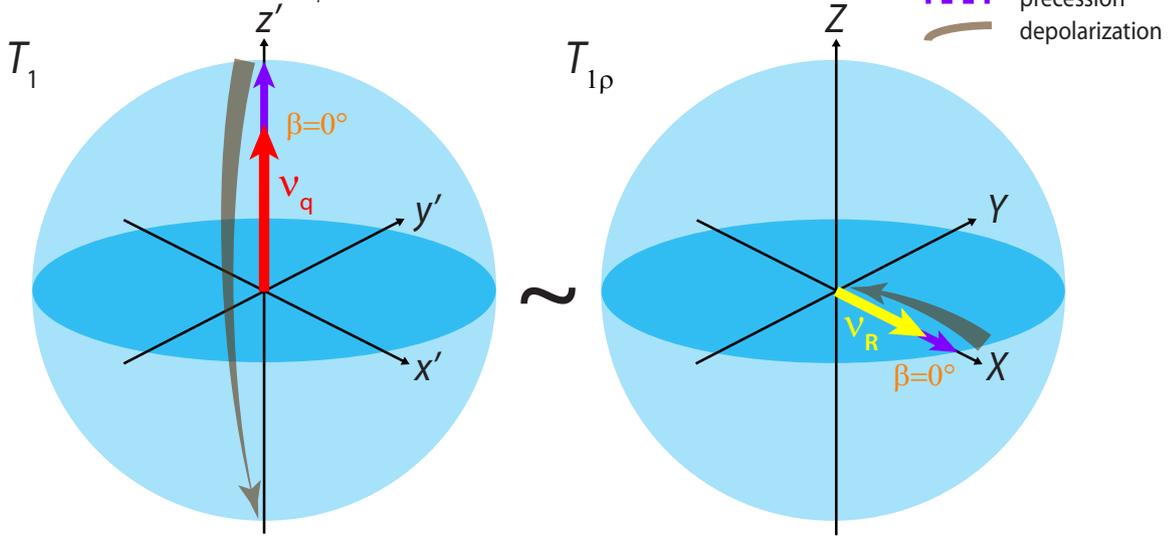
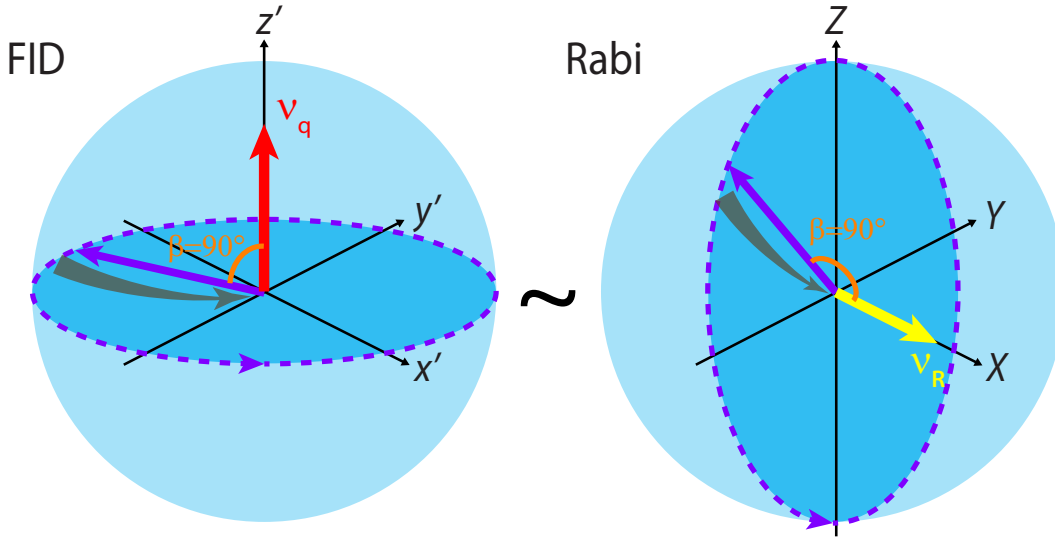

**Supplementary Figure S2. System dynamics in analogous experiments between free and driven evolution.** (a) $T_1$ versus $T_{1\rho}$. The figures represent the typical case when the initial qubit state is parallel with the quantizing field ($\beta = 0$). Depolarization shown is based on the condition, $\nu_R \ll k_B T/h \ll \nu_q$, so that, in the steady state, $\langle \hat{\sigma}_{z'} \rangle \approx -1$ for the $T_1$ process, and $\langle \hat{\sigma}_X \rangle \approx 0$ for the $T_{1\rho}$ process. (b) FID versus Rabi. The dashed line indicates the precession, while depolarization happens simultaneously. The process shown ignores the longitudinal relaxation.

|  | **Decoherence during free evolution** | **Decoherence during driven evolution** |
|---|---|---|
| Working frame | Qubit frame | Rotating frame |
| Quantization axis | $z'$-axis | $X$-axis |
| Nutation freq. | level splitting $\nu_\mathrm{q}$ | Rabi frequency $\nu_\mathrm{R}$ |
| | Longitudinal relaxation | |
| Method | Inversion recovery [26, 35] | Spin locking - $T_{1\rho}$ ([16], this work) |
| Process | $T_1 \ (= 1/\Gamma_1)$ | $T_{1\rho} \ (= 1/\Gamma_{1\rho})$ |
| Longitudinal decay | $z'$-axis | $X$-axis |
| Starting state | $|+z'\rangle$ | $|\pm X\rangle$ |
| Steady state | $|-z'\rangle$ | $|\langle \hat{\sigma}_X \rangle = 0\rangle$ |
| Decay rate | $\Gamma_1$ (exponential) | $\Gamma_{1\rho}$ (exp.) |
| Decay law | $\exp\{-\Gamma_1 \tau\}$, where $\Gamma_1 = \frac{1}{2} S_{x'}(\nu_\mathrm{q})$ | $\exp\{-(\frac{1}{2}\Gamma_1 + \Gamma_\nu)\tau\}$, where $\Gamma_\nu = \frac{1}{2} S_{z'}(\nu_\mathrm{R})$ |
| Noise susceptibility |  |  |
| • in general | $S_{\perp z'}(\nu_\mathrm{q})$ | $S_{\perp X}(\nu_\mathrm{R})$ |
| • flux qubit | $S_{x'}(\nu_\mathrm{q})$ | $S_{x'}(\nu_\mathrm{q} \pm \nu_\mathrm{R} \approx \nu_\mathrm{q}), \ S_{z'}(\nu_\mathrm{R})$ |
| Noise of interest | $S_{x'}(\nu_\mathrm{q})$ | $S_{z'}(\nu_\mathrm{R})$ |
| | Transverse decoherence | |
| Method | FID [26, 36], CPMG [12] | Rabi [12], Rotary echo [18] |
| Process | $T_2 \ (= 1/\Gamma_2)$ | $T_{2\rho} \ (= 1/\Gamma_{2\rho})$ |
| Transverse decay | $x'-y'$ plane | $Y-Z$ plane |
| Starting state | e.g., $|\pm x', \pm y'\rangle$ | $|-Z\rangle \ (= |-z'\rangle)$ |
| Steady state | $|\langle \hat{\sigma}_{x',y'} \rangle = 0\rangle$ | $|\langle \hat{\sigma}_{Y,Z} \rangle \approx 0\rangle$ |
| Decay rate |  |  |
| • Bloch-Redfield | $\Gamma_2 = \frac{1}{2}\Gamma_1 + \Gamma_\varphi$ (exp.) | $\Gamma_{2\rho} = \frac{1}{2}\Gamma_{1\rho} + \Gamma_{\varphi\rho}$ (exp.) |
| • $1/f$-type noise | exp.×Gaussian (linear) | exp.×Gaussian×algebraic (quadratic) |
| Decay law (FID/Rabi) | $\exp\{-\frac{1}{2}\Gamma_1\tau - \frac{(2\pi)^2}{2}\langle \delta\nu^2\rangle \tau^2\}$ $\times \cos(2\pi \Delta\nu \tau)$ | $\exp\{-(\frac{3}{4}\Gamma_1 + \frac{1}{2}\Gamma_\nu)\tau - \frac{(2\pi)^2}{2}\langle \delta\nu_\mathrm{R}^2\rangle \tau^2\}$ $\times [1 + (2\pi \frac{\langle \delta\nu^2\rangle}{\nu_\mathrm{R}}\tau)^2]^{-1/4} \cos(2\pi \nu_\mathrm{R} \tau)$ |
| Noise susceptibility |  |  |
| • in general | $\Gamma_1: S_{\perp z'}(\nu_\mathrm{q}); \ \Gamma_\varphi: \Delta\nu$ inhom. | $\Gamma_{1\rho}: S_{\perp X}(\nu_\mathrm{R}); \ \Gamma_{\varphi\rho}: \nu_\mathrm{R}'$ inhom. |
| • flux qubit | $\Gamma_1: S_{x'}(\nu_\mathrm{q});$ $\Gamma_\varphi: \Delta\nu$ inhom. | $\Gamma_{1\rho}: S_{x'}(\nu_\mathrm{q} \pm \nu_\mathrm{R} \approx \nu_\mathrm{q}), S_{z'}(\nu_\mathrm{R});$ $\Gamma_{\varphi\rho}: S_{x'}(\nu_\mathrm{q}), \nu_\mathrm{R}$ and $\Delta\nu$ inhom. |
| Noise of interest | $\langle \delta\nu^2 \rangle$ | $\Gamma_\nu = \frac{1}{2} S_{z'}(\nu_\mathrm{R})$ |

**Supplementary Table S**1. Comparison of decohering properties of a flux qubit during free evolution and driven evolution (weak and resonant).

# SUPPLEMENTARY NOTES

Our test device is a superconducting flux qubit, an aluminium loop interrupted by four Al-AlO$_x$-Al Josephson junctions. In the presence of external harmonic control, the Hamiltonian of the superconducting flux qubit in the laboratory frame (Supplementary Figure S1a) is

$$\hat{\mathcal{H}} = \frac{h}{2}\left[\Delta\hat{\sigma}_z + \varepsilon\hat{\sigma}_x + A_{\rm rf}\cos(2\pi\nu_{\rm rf}t + \phi)\hat{\sigma}_x\right], \tag{S1}$$

where $\hat{\sigma}_z$ and $\hat{\sigma}_x$ are Pauli operators. $\varepsilon = 2I_{\rm p}\Phi_{\rm b}/h$ ($I_{\rm p}$ is the superconducting persistent loop current) is called the energy flux bias, which gives the energy difference between the two classical circulating-current states (clockwise and counterclockwise) and is adjusted by the external magnetic flux $\Phi$ via $\Phi_{\rm b} = \Phi - \Phi_0/2$ ($\Phi_0$ is the superconducting flux quantum). $\Delta = 5.4\,\rm GHz$ is the tunnel-coupling strength between the two states. The oscillating term represents the harmonic drive with amplitude $A_{\rm rf}$, carrier frequency (or microwave frequency) $\nu_{\rm rf}$ and phase $\phi$. In this device, noise in $\varepsilon$, $\Delta$ and $A_{\rm rf}$ are considered the main physical sources of decoherence in most experiments, and have simple geometry in the lab frame. However, the decohering properties are routinely defined and better understood in the frame spanned by the qubit eigenbasis.

By making the transformation,

$$\begin{aligned}\hat{\sigma}_{x'} &= \cos\theta\,\hat{\sigma}_x - \sin\theta\,\hat{\sigma}_z\,, \\ \hat{\sigma}_{y'} &= \hat{\sigma}_y\,, \\ \hat{\sigma}_{z'} &= \cos\theta\,\hat{\sigma}_z + \sin\theta\,\hat{\sigma}_x\,,\end{aligned} \tag{S2}$$

where $\theta = \arctan(\varepsilon/\Delta)$ is the rotation angle of the quantization axis from the lab frame, we can write the Hamiltonian in Eq. S1 in the qubit (eigenbasis) frame (Supplementary Figure S1b):

$$\hat{\mathcal{H}}' = \frac{h}{2}\left[\nu_{\rm q}\hat{\sigma}_{z'} + A_{\rm rf}\cos\theta\cos(2\pi\nu_{\rm rf}t+\phi)\hat{\sigma}_{x'} + A_{\rm rf}\sin\theta\cos(2\pi\nu_{\rm rf}t+\phi)\hat{\sigma}_{z'}\right] \tag{S3}$$

where $\nu_{\rm q} = \sqrt{\varepsilon^2 + \Delta^2}$ is the qubit's level-splitting.

The free-evolution dynamics can be visualized on the Bloch sphere in this qubit frame (Supplementary Figure S2). Ignoring the external control, the only term left is the one along the quantization axis. The qubit, like a spin, undergoes right-handed (counterclockwise as viewed from the north pole) precession around the $z'$-axis, at a cycling rate equivalent to the qubit frequency. The qubit's Bloch vector and the $z'$-axis form an angle $\beta$ depending on the initial state. For example, if the qubit starts from the ground ($|-z'\rangle$, south pole) or excited state ($|+z'\rangle$, north pole), $\beta = 180°$ or $0°$. Hence the qubit state remains unchanged. When the qubit is initially prepared at a 50%-superposition state, e.g., $|\pm x'\rangle$ or $|\pm y'\rangle$, $\beta = 90°$ and the qubit will precess around the equatorial plane. In general, for an arbitrary value of $\beta$, the qubit leaves a cone-like trace which is axially symmetric around $z'$.

The free-evolution decohering processes are defined with respect to the quantization axis. Note that, decoherence in superconducting qubits should be understood in a sense of time-ordered ensemble average.

$T_1$ is the longitudinal depolarization (longitudinal relaxation time), or often just "relaxation time" for short. From perturbation theory, such longitudinal relaxation is the consequence of environmental modes at the qubit frequency transversely coupled to the qubit. The inversion recovery experiment [26], which initializes the qubit at the excited state and records that how it temporally relaxes along $z'$ (Supplementary Figure S2a), is one of the standard methods for measuring $T_1$.

$T_2$ is the transverse depolarization (transverse relaxation time), or just "coherence/dephasing time". In the weak coupling regime where decoherence of the qubit during the correlation time of the environment is negligible, the Bloch-Redfield theory [37–39] gives a simple description of the exponential dephasing rate: $1/T_2 = 1/2T_1 + 1/T_\varphi$, where $T_\varphi$ is the pure dephasing time. Unfortunately, in many instances such as $1/f$-type spectrum, the Bloch-Redfield approach does not apply because the noise correlation times become large at low frequency. However, the decay functions due to $T_1$ processes and due to $T_\phi$ processes still factorize [40]. Pure dephasing is a consequence of ensemble averaging over a fluctuating qubit frequency [41].

Depending on the experiment, there are various types of "$T_2''$". The simplest one is the free induction decay (FID) or Ramsey experiment [26], which measures the associated coherence time $T_{\rm FID}$ with bare precession (one-way phase accumulation). The measured pure dephasing is the result of ensemble averaging over a inhomogeneous parameters [14] (here, temporal inhomogeneity due to low-frequency noise for superconducting qubits). If a refocusing pulse is added in the middle of the free-evolution period as in the spin-echo experiment [36], dephasing due to low-frequency noise can be reduced. In superconducting qubits, the noise PSD is $1/f$-type, and we typically obtain a longer echo phase decay time $T_{\text{Echo}} > T_{\text{FID}}$. More advanced sequences such as CPMG [12] (a generalized version of the spin echo sequence) may further improve coherence and, in addition, be utilized to extract the free-evolution PSD through its noise-shaping/filtering properties.

Driven-evolution dynamics are conveniently described in a reference frame which rotates around $z'$ at the drive frequency $\nu_{\text{rf}}$ by a second transformation,

$$\begin{aligned}
\hat{\sigma}_X &= \cos(2\pi\nu_{\text{rf}}t)\hat{\sigma}_{x'} + \sin(2\pi\nu_{\text{rf}}t)\hat{\sigma}_{y'}\ , \\
\hat{\sigma}_Y &= \cos(2\pi\nu_{\text{rf}}t)\hat{\sigma}_{y'} - \sin(2\pi\nu_{\text{rf}}t)\hat{\sigma}_{x'}\ , \\
\hat{\sigma}_Z &= \hat{\sigma}_{z'}\ .
\end{aligned} \quad (S4)$$

The Hamiltonian in Eq. S3 is then written as (Supplementary Figure S1c)

$$\begin{aligned}
\hat{\tilde{\mathcal{H}}} = \frac{h}{2}\Big[&\Delta\nu\hat{\sigma}_Z + \frac{1}{2}A_{\text{rf}}\cos\theta\big(\cos\phi\,\hat{\sigma}_X + \sin\phi\,\hat{\sigma}_Y\big) \\
&+ \frac{1}{2}A_{\text{rf}}\cos\theta\big(\cos(-2\pi\cdot 2\nu_{\text{rf}}t - \phi)\hat{\sigma}_X + \sin(-2\pi\cdot 2\nu_{\text{rf}}t - \phi)\hat{\sigma}_Y\big) \\
&+ A_{\text{rf}}\sin\theta\cos(2\pi\nu_{\text{rf}}t + \phi)\hat{\sigma}_Z\Big]\ ,
\end{aligned} \quad (S5)$$

where $\Delta\nu = \nu_{\text{q}} - \nu_{\text{rf}}$ is the frequency detuning between the qubit and the driving field. In the weak driving limit ($A_{\text{rf}} \ll \nu_{\text{rf}}$), the last two lines in Eq. S5 can be omitted, since these rapid oscillations average to zero on any appreciable time scale of the qubit dynamics in the rotating frame (rotating wave approximation). The Hamiltonian within this approximation is given by

$$\hat{\tilde{\mathcal{H}}} = (h/2)\big[\Delta\nu\hat{\sigma}_Z + \nu_{\text{R}}(\cos\phi\,\hat{\sigma}_X + \sin\phi\,\hat{\sigma}_Y)\big]\ , \quad (S6)$$

where $\nu_{\text{R}} = \frac{1}{2}A_{\text{rf}}\cos\theta$ is the Rabi frequency under resonant driving ($\nu_{\text{rf}} = \nu_{\text{q}}$). In the presence of finite detuning, Eq. S6 describes the effective driving field, whose effective Rabi frequency, $\nu'_{\text{R}} = \nu_{\text{R}}\sqrt{1 + (\Delta\nu/\nu_{\text{R}})^2} \approx \nu_{\text{R}} + \Delta\nu^2/2\nu_{\text{R}}$ is linear with $\nu_{\text{R}}$ and quadratic with $\Delta\nu$.

For resonant driving along the $X$-axis ($\Delta\nu = 0$, $\phi = 0$), the Hamiltonian in Eq. S6 describes a pseudospin quantized along $X$ with a level-splitting equivalent to the Rabi frequency $\nu_{\text{R}}$ (Supplementary Figure S1d). Therefore, the driven-evolution dynamics represented in the rotating frame are analogous to the free-evolution ones in the qubit frame. That is, the pseudospin precesses around the $X$ axis at the rate of $\nu_{\text{R}}$, and its subtended angle depends on the initial state. For example, if the qubit is prepared at either $|+X\rangle$ or $|-X\rangle$ state, its Bloch vector will remain in that orientation, as does the corresponding free-evolution scenario when the qubit is initialized in its excited ($|+z'\rangle$) or ground ($|-z'\rangle$) state. When the initial state is an "equatorial state" (here, lying on the $Y-Z$ plane), the pseudospin will precess around the $Y-Z$ plane just as their free-evolution counterparts precess around the $x'-y'$ plane (equatorial states for the free-evolution case).

Following this analogy, we define decohering processes for the pseudospin with respect to the new quantization axis $X$. $T_{1\rho}$ is the rotating-frame longitudinal depolarization/relaxation time, i.e. relaxation in the rotating frame, which corresponds to depolarization along the $X$-axis. The environmental noise responsible for this decay is the transversely coupled ($\perp X$, comprising fluctuations $\nu_{\text{R}}$ of the Rabi-frequency). To measure $T_{1\rho}$, we use the $T_{1\rho}$ experiment described in this manuscript, which prepares the qubit at the $|\pm X\rangle$ state, and then records how it depolarizes during continuous driving.

$T_{2\rho}$ is the rotating-frame transverse depolarization/relaxation time. Similar to the free-evolution case, this transverse decoherence can also be factorized into a $T_{1\rho}$-induced part and a pure dephasing part $T_{\varphi\rho}$, all with respect to the new quantization axis. The rotating-frame pure dephasing is a consequence of fluctuation in the pseudospin frequency. It has first-order sensitivity to noise in the Rabi frequency $\nu_{\text{R}}$ (due to driving field amplitude fluctuations), which is found to be the dominating source of Rabi decay in this device [18].

$T_{2\rho}$, like its free-evolution counterpart, also varies in different experiments. The simplest one is the original Rabi experiment [26], which measures the decaying oscillatory signal (characteristic time $T_{\text{Rabi}}$) by resonantly driving the qubit from its ground state (the Rabi experiment in a general sense means the driven evolution starting from an arbitrary state. Hence, even the $T_{1\rho}$ experiment is just a special case of Rabi. However, in this article, we denote Rabi as its most original version only, in which the qubit starts from its ground state $|-z'\rangle$ ($=|-Z\rangle$)). It is the driven-evolution analog of the FID experiment, since both characterize the transverse decay during "bare" equatorial precession. The pure dephasing part is sensitive to $\nu_{\text{R}}$ (linear coupling, first order) and $\Delta\nu$ (quadratic coupling, second order) inhomogeneities, both of which modify the pseudospin frequency, i.e. the effective Rabi frequency $\nu'_\text{R}$. If the second half of the Rabi pulse is 180°-phase-shifted as in the so-called rotary-echo experiment [18], the Rabi field in the rotating frame points to $-X$ instead of $X$, and the precession reverses direction. Therefore, slow fluctuations in the pseudospin frequency can be refocused. Rotary-echo is the driven-evolution analog of spin-echo, and the decay (characteristic time $T_\text{Rotary}$) can be analyzed in a similar way. Furthermore, the analogy to the CPMG sequence is a multi-pulse generalization of the rotary-echo sequence.

The connection between the driven-evolution times ($T_{1\rho}$, $T_{2\rho}$), the free-evolution times ($T_1$, $T_2$), and the power spectral density (PSD) of the physical noise terms (noise in $\Delta$ and $\varepsilon$) is formally made using the generalized Bloch equations (GBE) [23, 24] and the transformations between Eqs. S1-S6.

The qubit-bath interaction Hamiltonian is defined as

$$\hat{\mathcal{H}}_\text{I} = \frac{h}{2}\big[\hat{Q}_\Delta\,\hat{\sigma}_z + \hat{Q}_\varepsilon\,\hat{\sigma}_x\big]\,, \tag{S7}$$

where $\hat{Q}_\Delta$ and $\hat{Q}_\varepsilon$ represent the bath variables which couple to the effective tunnel-coupling and flux terms, respectively.

When transformed into the qubit frame, the Hamiltonian becomes

$$\hat{\mathcal{H}}'_\text{I} = \frac{h}{2}\big[(\hat{Q}_\Delta \cos\theta + \hat{Q}_\varepsilon \sin\theta)\hat{\sigma}_{z'} + (-\hat{Q}_\Delta \sin\theta + \hat{Q}_\varepsilon \cos\theta)\hat{\sigma}_{x'}\big]\,. \tag{S8}$$

Therefore, the PSD in the qubit frame is a (co)sinusoidally weighted sum of the PSD in the lab frame, namely,

$$\begin{aligned} S_{x'}(f) &= \sin^2\theta\, S_\Delta(f) + \cos^2\theta\, S_\varepsilon(f)\,, \\ S_{z'}(f) &= \cos^2\theta\, S_\Delta(f) + \sin^2\theta\, S_\varepsilon(f)\,. \end{aligned} \tag{S9}$$

$S_\lambda(f)$ stands for the PSD of $\lambda$-noise. It is defined as the reduced power spectral density, i.e., $S_\lambda(f) = (2\pi)^2 \int_{-\infty}^\infty dt \frac{1}{2}\langle \lambda(0)\lambda(t) + \lambda(t)\lambda(0)\rangle \exp(-i2\pi ft)$ and $\frac{1}{2}\langle \lambda(0)\lambda(t) + \lambda(t)\lambda(0)\rangle$ is the symmetrized autocorrelation function. In the classical regime ($hf < k_\text{B}T$, where $T$ is the effective temperature), it can be replaced by the classical autocorrelation function, $\langle \lambda(0)\lambda(t)\rangle$. $\lambda \in \{x', y', z'\ldots$ or $\Delta(=\hat{Q}_\Delta), \varepsilon(=\hat{Q}_\varepsilon)\ldots\}$ indicates either the fluctuating part of system Hamiltonian or the axis along which those fluctuations occur.

In the rotating frame, the stochastic Hamiltonian becomes

$$\begin{aligned} \hat{\tilde{\mathcal{H}}}_\text{I} = \frac{h}{2}\big[&(\hat{Q}_\Delta \cos\theta + \hat{Q}_\varepsilon \sin\theta)\hat{\sigma}_Z + \ldots \\ &+(-\hat{Q}_\Delta \sin\theta + \hat{Q}_\varepsilon \cos\theta)\cos(2\pi\nu_\text{rf}t)\hat{\sigma}_X - \ldots \\ &-(-\hat{Q}_\Delta \sin\theta + \hat{Q}_\varepsilon \cos\theta)\sin(2\pi\nu_\text{rf}t)\hat{\sigma}_Y\big]\,. \end{aligned} \tag{S10}$$

This Hamiltonian, together with the deterministic one in Eq. S6, governs the system's motion for a single realization. Comparing Eq. S8 and Eq. S10, $z'$-noise in the qubit frame is invariably transferred into $Z$-noise, while $x'$-noise is separated into $X$-noise and $Y$-noise with a frequency mixing of $\nu_\text{rf}$, indicating that the rotating-frame noise $S_{X,Y}(f)$ is actually modulated (frequency-shifted) qubit-frame noise, derived from $S_{x'}(|f \pm \nu_\text{rf}|)$.

In the weak coupling regime, ensemble averaging over all the heat bath realizations gives the non-Markovian equations of motion, from which the GBE are derived. With resonant driving and rotating wave approximation, we have the simplified GBE:

$$\begin{aligned} \frac{d\langle \hat{\sigma}_X\rangle}{dt} &= -\Gamma_X \langle \hat{\sigma}_X\rangle + v_X\,, \\ \frac{d\langle \hat{\sigma}_Y\rangle}{dt} &= -\Gamma_Y \langle \hat{\sigma}_Y\rangle + \nu_\text{R}\langle \hat{\sigma}_Z\rangle + v_Y\,, \\ \frac{d\langle \hat{\sigma}_Z\rangle}{dt} &= -\Gamma_Z \langle \hat{\sigma}_Z\rangle - \nu_\text{R}\langle \hat{\sigma}_Y\rangle + v_Z\,. \end{aligned} \tag{S11}$$

The depolarization coefficients are determined by the PSDs of the heat bath fluctuations:

$$\begin{aligned} \Gamma_X &= \frac{1}{8}\big[S_{x'}(\nu_\text{q} + \nu_\text{R}) + S_{x'}(\nu_\text{q} - \nu_\text{R})\big] + \frac{1}{2}S_{z'}(\nu_\text{R})\,, \\ \Gamma_Y &= \frac{1}{4}S_{x'}(\nu_\text{q}) + \frac{1}{2}S_{z'}(\nu_\text{R})\,, \\ \Gamma_Z &= \frac{1}{4}S_{x'}(\nu_\text{q}) + \frac{1}{8}\big[S_{x'}(\nu_\text{q} + \nu_\text{R}) + S_{x'}(\nu_\text{q} - \nu_\text{R})\big]\,, \end{aligned} \tag{S12}$$

and $v_\lambda$ ($\lambda = X, Y, Z$) (see [23, 24] for details) defines the steady-state polarization:

$$\langle \hat{\sigma}_X \rangle^{\text{ss}} = \frac{v_X}{\Gamma_X} ,$$
$$\langle \hat{\sigma}_Y \rangle^{\text{ss}} = \frac{\Gamma_Z v_Y + \nu_{\text{R}} v_Z}{\nu_{\text{R}}^2 + \Gamma_Y \Gamma_Z} ,$$
$$\langle \hat{\sigma}_Z \rangle^{\text{ss}} = \frac{\Gamma_Y v_Z - \nu_{\text{R}} v_Y}{\nu_{\text{R}}^2 + \Gamma_Y \Gamma_Z} . \tag{S13}$$

If the system-environment coupling is weak enough such that the driven-evolution depolarization is negligible on the time scale of qubit precession in the rotating frame, i.e. $\nu_{\text{R}} \gg \Gamma_\lambda$ ($\lambda = X, Y, Z$), we have $\langle \hat{\sigma}_Y \rangle^{\text{ss}} = \langle \hat{\sigma}_Z \rangle^{\text{ss}} = 0$. In addition, $\langle \hat{\sigma}_X \rangle^{\text{ss}} = 0$, if $\nu_{\text{R}} \ll k_{\text{B}} T/h$.

When there is finite detuning ($\Delta\nu \neq 0$), the quantization axis of the pseudospin becomes the effective Rabi field, i.e. the pink arrow in Supplementary Figure S1c. In this case, the longitudinal steady-state polarization is expressed by

$$\langle \hat{\sigma}_{Z'} \rangle^{\text{ss}} = \frac{\sin\eta\, S_{x'}(\nu_{\text{q}}) \tanh(\frac{h\nu_{\text{q}}}{2k_{\text{B}}T}) + \cos^2\eta\, S_{z'}(\nu_{\text{R}}) \tanh(\frac{h\nu_{\text{R}}}{2k_{\text{B}}T})}{\frac{1}{2}(1+\sin^2\eta)\, S_{x'}(\nu_{\text{q}}) + \cos^2\eta\, S_{z'}(\nu_{\text{R}})} , \tag{S14}$$

where $Z'$ indicates the longitudinal axis of the pseudospin in the presence of small detuning ($\Delta\nu < \nu_{\text{R}}$), and $\eta = \arctan(\Delta\nu/\nu_{\text{R}})$ is its tilted angle from the equatorial plane. In both the weak driving and weak coupling (driven-evolution) limits, and to first order in detuning (or $\eta$), the steady-state $X$-polarization is

$$\langle \hat{\sigma}_X \rangle^{\text{ss}} = \cos\eta \langle \hat{\sigma}_{Z'} \rangle^{\text{ss}} \approx -\frac{\sin\eta\, S_{x'}(\nu_{\text{q}})}{\frac{1}{2} S_{x'}(\nu_{\text{q}}) + S_{z'}(\nu_{\text{R}})} . \tag{S15}$$

The component $\langle \hat{\sigma}_{X'} \rangle^{\text{ss}}$ is negligible in the weak coupling limit.

Connecting the results in Eq. S12 with the $T_{1\rho}$ and Rabi experiments, we have

$$\Gamma_{1\rho} = \Gamma_X \quad \text{and} \quad \Gamma_{1\rho}^\dagger \equiv \Gamma_{\text{Rabi}}^\dagger = \frac{1}{2}(\Gamma_Y + \Gamma_Z) = \frac{1}{2}\Gamma_{1\rho} + \Gamma_{\varphi\rho} , \tag{S16}$$

where $\Gamma_{\varphi\rho} = \frac{1}{4} S_{x'}(\nu_{\text{q}})$ is the rotating-frame pure dephasing rate in the Rabi experiment. Note that, in this model, we did not take into account the low-frequency fluctuations in $\Delta\nu$ and $\nu_{\text{R}}$, which, in fact, dominate the Rabi decay. Therefore, we use the "†" symbol to indicate that the rate is free from those inhomogeneities. However, decay due to $\Delta\nu$ and $\nu_{\text{R}}$ inhomogeneities can be calculated semiclassically, in the same way as is done in the FID case [41].

As previously discussed, the longitudinal relaxation rate $\Gamma_{1\rho}$ depends on the transversely coupled (rotating-frame) noise at the frequency of $\nu_{\text{R}}$, i.e. $S_Z(\nu_{\text{R}})$ and $S_Y(\nu_{\text{R}})$. While $S_Z(\nu_{\text{R}})$ is equivalent to $S_{z'}(\nu_{\text{R}})$, $S_Y(\nu_{\text{R}})$ is derived from $S_{x'}(\nu_{\text{q}} \pm \nu_{\text{R}})$ due to the frequency mixing during frame transformation. This explains all the PSD components that contribute to $\Gamma_{1\rho}$. On the other hand, the transverse decoherence $\Gamma_{\text{Rabi}}^\dagger$ combines a relaxation-induced part $\frac{1}{2}\Gamma_{1\rho}$ and pure dephasing $\Gamma_{\varphi\rho}$ in the rotating frame. The presence of the qubit-frequency noise $S_{x'}(\nu_{\text{q}})$ in the expression for $\Gamma_{\varphi\rho}$ is due to the fact that, when transferred into the rotating frame, $S_{x'}(\nu_{\text{q}})$ is mixed down to zero-frequency noise, $S_X(0)$, which acts as the longitudinal quasistatic noise leading to dephasing. From above, we see that the decoherence formalism of driven evolution in the rotating frame is truly analogous to that of free evolution in the qubit frame.

At weak driving, we have $\nu_{\text{q}} \pm \nu_{\text{R}} \approx \nu_{\text{q}}$, which further simplifies Eq. S16:

$$\Gamma_{1\rho} = \frac{1}{2}\Gamma_1 + \Gamma_\nu ,$$
$$\Gamma_{\text{Rabi}}^\dagger = \frac{3}{4}\Gamma_1 + \frac{1}{2}\Gamma_\nu , \tag{S17}$$

where

$$\Gamma_1 = \frac{1}{2} S_{x'}(\nu_{\text{q}}) = \frac{1}{2}\left[\sin^2\theta\, S_\Delta(\nu_{\text{q}}) + \cos^2\theta\, S_\varepsilon(\nu_{\text{q}})\right] ,$$
$$\Gamma_\nu = \frac{1}{2} S_{z'}(\nu_{\text{R}}) = \frac{1}{2}\left[\cos^2\theta\, S_\Delta(\nu_{\text{R}}) + \sin^2\theta\, S_\varepsilon(\nu_{\text{R}})\right] . \tag{S18}$$

$\Gamma_1$ is the free-evolution longitudinal relaxation rate, and $\Gamma_\nu$ is the rate associated with the Rabi-frequency noise. Their relation to $\Gamma_{1\rho}$ inspires us to extract noise PSD at the Rabi frequency by measuring the lab-frame and rotating-frame longitudinal relaxation rates and taking the subtraction.

---